\documentclass[11pt]{article}
\usepackage{graphicx,ctable}
\begin{document}
\begin{center}
{\Large\bf{Thermal Model Description of p--Pb Collisions at $\sqrt{s_{NN}} $  = 5.02 TeV }}\\
\vspace*{1cm}  
{\sf{Natasha Sharma$^1${\footnote{email: natasha.sharma@cern.ch}}, Jean Cleymans$^2${\footnote{email: jean.cleymans@uct.ac.za}}, 
Lokesh Kumar$^1${\footnote{email: kumar.lokesh@cern.ch}}}},\\[0.1cm]
\vspace{10pt}  
{\small  {\em 
$^1$ Department of Physics, Panjab University, Chandigarh 160014, India.\\
$^2$UCT-CERN Research Centre and Department of Physics, University of Cape Town,\\ Rondebosch 7701, South Africa.}}\\
\normalsize  
\end{center}

\begin{center}
{\large In Memory of Helmut Oeschler}
\end{center}
\date{\today}

\begin{abstract}
The ALICE data on light flavor hadron production obtained in $p-Pb$ collisions at $\sqrt{s_{NN}} $  = 5.02 TeV  
are studied in the thermal model  using the canonical approach with exact strangeness conservation. 
The chemical freeze-out temperature is independent of centrality 
except for the lowest multiplicity bin,  
with values close to 160 MeV but
 consistent  with those obtained in $Pb-Pb$ collisions
at $\sqrt{s_{NN}}$ = 2.76 TeV. 
The value of the strangeness non-equilibrium factor $\gamma_s$ is slowly increasing with multiplicity from
0.9 to 0.96, i.e. it is always very close to full chemical equilibrium. 
\end{abstract}


\vspace{1mm}
%%%%%%%%%%%%%%%%%%%%%%%%%%%%%%%%%%%%%%%%%%%%%%%%%%%%%%%%%%%%%%%%%%%%%%%%%%%%%%%%%%%%%%%%%%%%%%%%%%%%%%%
\section{Introduction}
%%%%%%%%%%%%%%%%%%%%%%%%%%%%%%%%%%%%%%%%%%%%%%%%%%%%%%%%%%%%%%%%%%%%%%%%%%%%%%%%%%%%%%%%%%%%%%%%%%%%%%%
Over the last three decades, the hadron resonance gas (thermal model for short) in its grand-canonical (GC)
and canonical formulations has 
been very successful in describing the abundances of light flavored hadron obtained in heavy-ion 
collisions~\cite{Lorenz:2017hjp,Adamczyk:2017iwn,Floris:2016hhc} up to the highest beam energies. This model uses a minimal number
of parameters, mainly, the chemical freeze-out temperature $T_{ch}$, the baryon chemical potential $\mu_B$ and the volume $V$.
Some questions about this approach have been raised recently, e.g.:\\
- what is the effect of an incomplete list of hadronic resonances~\cite{Bazavov:2014xya,Alba:2017mqu,Alba:2014eba, Bluhm:2013yga}?\\
- are there  separate freeze-out temperature for strange particles~\cite{Chatterjee:2016cog,Bellwied:2017uat}?\\
- is the theoretical description complete~\cite{Vovchenko:2016rkn}?\\

The first question has to be answered by new experimental results on hadronic resonances.
The latter two cases naturally lead to the introduction of new extra parameters which reduces the simplicity of the model.

In the present paper we 
do not address these questions but 
investigate in detail the particle abundances in $p-Pb$  collisions~\cite{Abelev:2013haa,Adam:2015vsf,Adam:2016bpr,Vislavicius:2016rwi}
in the hope that  this contributes to the understanding and  the status of the model description.

At LHC energies, the particle abundances in  central heavy-ion collisions have been well
described~\cite{Floris:2016hhc, Andronic:2017pug,Becattini:2017pxe} 
over nine orders of magnitude with only two parameters $T_{ch}$ and $V$, with $\mu_B$ being restricted to 
zero because
of the particle-antiparticle symmetry at the LHC.

Our results show that there is no clear  dependence on the centrality, as measured by the values of   $dN_{ch}/d\eta$, for the 
chemical freeze-out temperature. This confirms results obtained earlier on 
the dependence of thermal parameters on the size of the system obtained~\cite{Cleymans:2001at}.

In section II we briefly review the main features of the model. Section III presents our results for $p-Pb$ collisions.
Section IV contains a discussion of results and Section V presents our conclusions.
\vspace*{1cm}

\section{The model}

In general, if the number of particles carrying quantum numbers related to a conservation law is small, then
the grand-canonical description no longer holds. In such a case the conservation of quantum numbers has to be implemented
exactly in the canonical ensemble~\cite{Hagedorn:1984uy,Cleymans:1990mn}.
%\cite{BraunMunzinger:2001as,BraunMunzinger:2003zd}. 
Here, we refer only  to strangeness conservation and
consider charge and baryon number conservation to be fulfilled on the average in the grand canonical
ensemble because the number of charged particles and baryons is much larger than that of strange
particles~\cite{BraunMunzinger:2001as}.

%We start by presenting a brief reminder of the general concepts of the %thermal model.
%
%\subsection{Grand Canonical Ensemble}
%
In the {\it Grand-Canonical Ensemble (GCE)} , the volume $V$, temperature $T_{ch}$ and the  chemical 
potentials $\vec \mu$ 
determine the partition function $Z(T,V,\vec\mu)$. In the hadronic fireball of non-interacting 
hadrons, $\ln Z$ is the sum of the contributions of all $i$-particle species 
\begin{equation}
\frac{1}{V} \, {\rm ln}Z(T, V, \vec{\mu}) = \sum_i {}Z_i^1(T, {\vec\mu}), \label{equ1}
\end{equation}
where  $\vec{\mu}=(\mu_B,\mu_S,\mu_Q)$ are
the chemical potentials related to the conservation of baryon number, strangeness and electric charge, respectively. 
$Z^1_i$ is the one-particle partition function.

The partition function  contains all information needed to obtain the number density $n_i$ of 
particle species $i$. Introducing the particle's specific chemical potential $\mu_i$, one gets
\begin{equation}
n_i^{}(T, \vec{\mu}) = \frac{1}{V}\left. \frac{\partial(T \ln Z)}{\partial\mu_i}\right|_{\mu_i=0}. \label{equ3}
\end{equation}
Any resonance that decays into species $i$  contributes to the yields eventually measured. Therefore, 
the contributions from all heavier hadrons that decay to hadron $i$ 
are included.

%with the branching fraction 
%$\Gamma_{j \rightarrow i}$ are
%
%\begin{equation}
%n_i^{\rm decay} = \sum_j  \Gamma_{j \rightarrow i} ~ n_j. \label{equ4}
%\end{equation}
%
%Consequently, the final yield $N_i$ of particle species $i$ is the sum of %the thermally
%produced particles and the decay products of resonances,
%%
%\begin{equation}
%N_i = (n_i^{} + n_i^{\rm decay}) ~ V. \label{equ5}
%\end{equation}
%
%From Eqs.~(\ref{equ3}-\ref{equ5}) it is clear that 
In  the GCE the particle yields are determined by the
volume of the fireball, its temperature and the chemical potentials.

%\subsection{Canonical ensemble}
%
If the number of particles  is small, conservation laws have to be implemented
exactly. % using the {\it Canonical Ensemble (CE)}. 
We refer here only to the {\it Strangeness Canonical Ensemble (SCE)} in which  strangeness conservation is considered and 
charge and baryon number are conserved  on the average. 
The density of strange particle $i$ carrying strangeness $s$ can be obtained from~\cite{BraunMunzinger:2001as},
\begin{eqnarray}
n_{i}^C&=&{{Z^1_{i}}\over {Z_{S=0}^C}} \sum_{k=-\infty}^{\infty}\sum_{p=-\infty}^{\infty} a_{3}^{p} a_{2}^{k}
a_{1}^{{-2k-3p- s}}\nonumber \\
&&I_k(x_2) I_p(x_3) I_{-2k-3p- s}(x_1),   \label{equ6}
\end{eqnarray}
where $Z^C_{S=0}$ is the canonical partition function
\begin{eqnarray}
Z^C_{S=0}&=&e^{S_0} \sum_{k=-\infty}^{\infty}\sum_{p=-\infty}^{\infty} a_{3}^{p}
a_{2}^{k} a_{1}^{{-2k-3p}}\nonumber 
 I_k(x_2) I_p(x_3) I_{-2k-3p}(x_1),
\label{eq7}
\end{eqnarray}
where $Z^1_i$ is the one-particle partition function calculated for $\mu_S=0$ in the Boltzmann
approximation. The arguments of the Bessel functions $I_s(x)$ and the parameters $a_i$ are introduced as,
\begin{eqnarray} a_s= \sqrt{{S_s}/{S_{\mathrm{-s}}}}~~,~~ x_s = 2V\sqrt{S_sS_{\mathrm{-s}}} \label{eq8a}, \end{eqnarray}
where $S_s$ is the sum  of all $Z^1_k(\mu_S=0)$  for particle species $k$ carrying strangeness
$s$. 

In the limit where $x_n<1$ (for $n=1$, 2 and 3)   the density of strange particles carrying
strangeness $s$ is well approximated by~\cite{BraunMunzinger:2001as}
\begin{equation}
%n_i^{SCE} \simeq n_i^{GCE} \frac{I_{s}(x_1)}{I_0(x_1)}. \label{equ7}
\frac{n_i^{SCE}} {n_i^{GCE} }\simeq  \frac{I_{s}(x_1)}{I_0(x_1)}. \label{equ7}
\end{equation}
From these  equations it is clear that in the canonical ensemble the strange particle density depends explicitly on the 
volume  through the arguments of the Bessel functions. 
It has been suggested that this volume might be different from the overall volume $V$~\cite{Cleymans:1998yb,Hamieh:2000tk}. 
In our analysis we will not entertain this possibility and work instead with an overall parameter $\gamma_s$
to describe the deviation of strange particles from chemical equilibrium~\cite{Letessier:1993hi}.
This amounts to replacing each particle density by
\begin{equation}
n_i^C \rightarrow \gamma_s^{|s|}n_i^C
\end{equation}
where $|s|$ is the sum of the number of strange and anti-strange quarks.
Note that the $\phi$ meson picks up a factor $\gamma_s^2$ since it contains a strange and an anti-strange quark.  This makes the $\phi$
meson behave in a similar way to the $\Xi$ baryon which contains two strange quarks, a behavior which is supported by the
data.  The changes for the other strange mesons and baryons are as follows:
\begin{eqnarray}
n_K^C &\rightarrow& \gamma_s n_K^C , \nonumber\\
n_{\Lambda}^C &\rightarrow& \gamma_s n_{\Lambda}^C  ,\nonumber\\
n_{\Xi}^C &\rightarrow& \gamma_s^2 n_{\Xi}^C ,\nonumber\\
n_{\Omega}^C &\rightarrow& \gamma_s^3 n_{\Omega}^C .\nonumber
\end{eqnarray}
%%%%%%%%%%%%%%%%%%%%%%%%%%%%%%%%%%%%%%%%%%%%%%%%%%%%%%%%%%%%%%%%%%%%%
\section{Scaling of yields with strange quark content}
\begin{figure}[hbt]
  % \begin{center}
 \includegraphics[trim=0cm 0cm 0cm 0cm, clip=true, width = 0.5\textwidth]{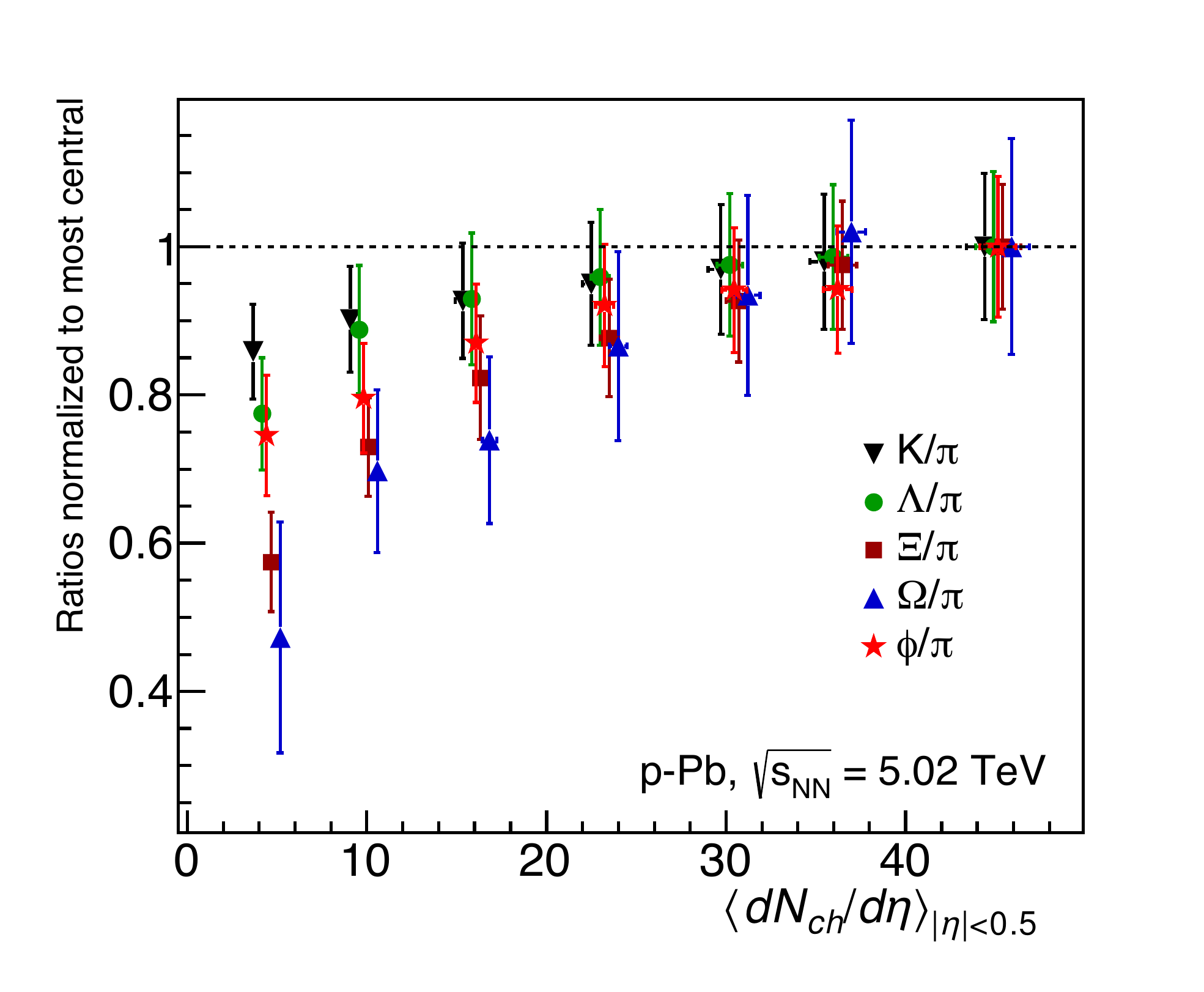} 
\includegraphics[trim=0cm 0cm 0cm 0cm, clip=true, width = 0.5\textwidth]{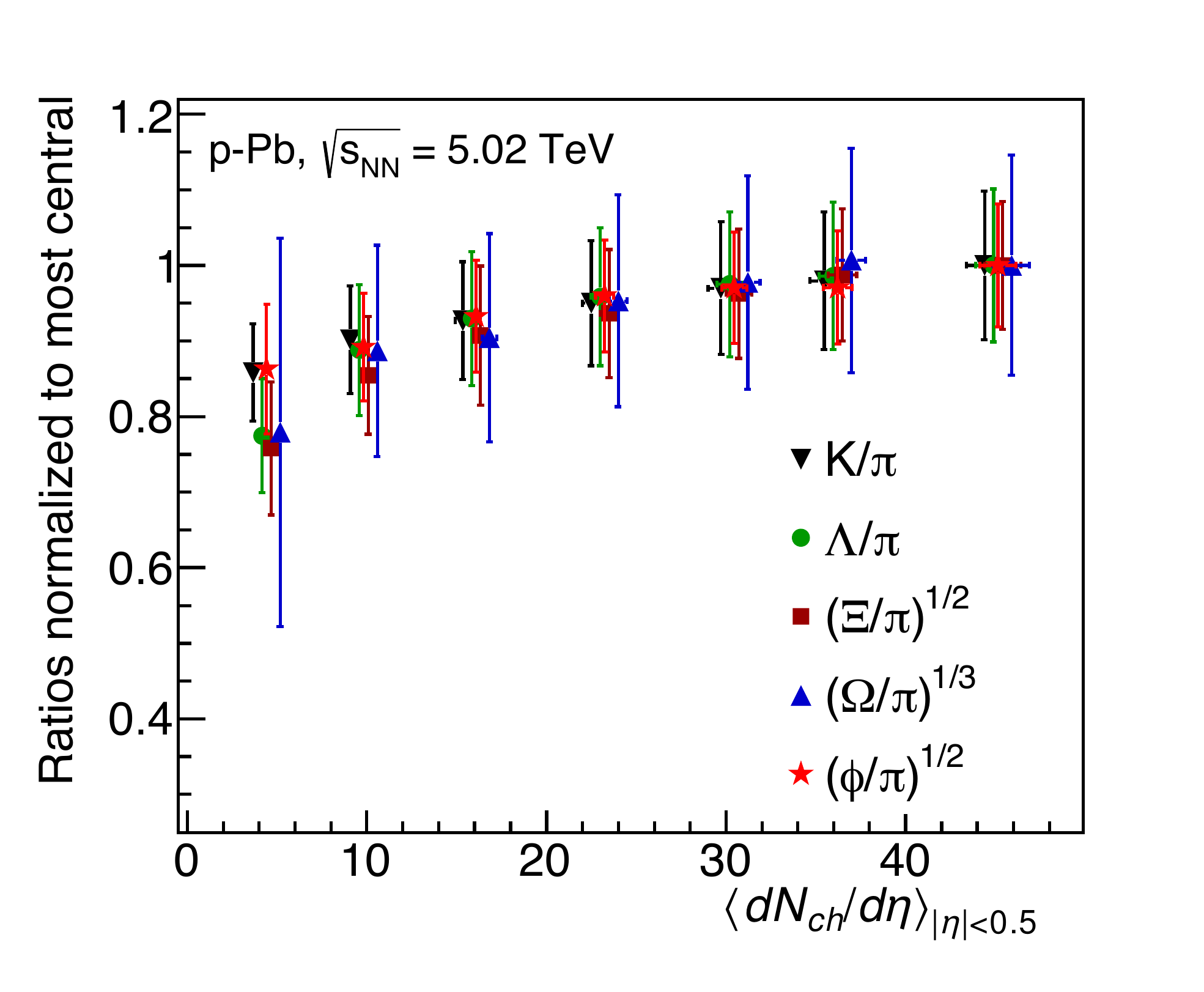} 
\caption{The ratios of strange particle yields to  pion yields, normalized to the corresponding value at the highest multiplicity, as 
a function of $\langle dN_{ch}/d\eta\rangle _{|\eta|<0.5}$ for p--Pb collisions at 5.02 TeV, left as measured, right scaled with a power
determined by the number of strange and antistrange quarks in the hadron.} 
    \label{fig:pPb-scaling}
%  \end{center}
\end{figure}

%%%%%%%%%%%%%%%%%%%%%%%%%%%%%%%%%%%%%%%%%%%%%%%%%%%%%%%%%%%%%%%%%%%%%
The experimental  results were taken from~\cite{Abelev:2013haa,Adam:2015vsf,Adam:2016bpr}, these are obtained for central rapidity in an 
interval $\Delta y$ = 1. 
One of the basic features of the SC approach is that the particle ratios exhibit an approximate scaling 
with the difference of the strangeness quantum numbers.   The support for this scaling 
is shown in  Fig.~\ref{fig:pPb-scaling}.  
The left pane shows the  ratios of various strange particle to pion yields i
%with $S$ varying from 1 to 3 divided by the pion yield, 
as a function of multiplicity. 
Following references~\cite{Abelev:2013haa,Adam:2015vsf,Adam:2016bpr} we use $K/\pi$ as an abbreviation for the ratio  $(K^++K^-)/(\pi^++\pi^-)$
and similarly for all other ratios, except that $\phi/\pi$ is an abbreviation for $\phi/(\pi^++\pi^-)/2$.
Scaling these ratios with the  power of the corresponding number of strange quarks
%$\Delta S$.
one observes a common trend (right pane of Fig.~\ref{fig:pPb-scaling}). 
 As can be seen from Fig. 1  scaling with the number of strange quarks is very well supported by the  experimental data.
This scaling  behavior with strange quark content is motivated by  the SCE description including the factor $\gamma_s$.
In particular, the yield of $\phi$ mesons closely follows the one of e.g. $\Xi$ baryons. This 
argues against the use of a different correlation 
radius for the strangeness suppression as this would not affect the $\phi$ meson since it has strangeness zero.
%
%\newpage
%%%%%%%%%%%%%%%%%%%%%%%%%%%%%%%%%%%%%%%%%%%%%%%%%%%%%%%%%%%%%%%%%%%%%
\section{Fits using the thermal model}
%%%%%%%%%%%%%%%%%%%%%%%%%%%%%%%%%%%%%%%%%%%%%%%%%%%%%%%%%%%%%%%%%%%%%
%
%
In this section we present fits to the hadronic yields
obtained in p-Pb collisions at $\sqrt{s_{NN}}$ = 5.02 TeV in seven multiplicity bins
such as 0-5\%, 10-20\%, 20-40\%, 40-60\%, 60-80\% and 80-100\%.
%, starting with the most central collisions.

We have used THERMUS~\cite{Wheaton:2004qb}  to perform these fits in the strangeness canonical ensemble (SCE) with 
$\mu_B$ and $\mu_Q$ fixed to 0 and using only one radius $R$ for 
the system.
The fits to the bins with the most central collisions are shown in Fig.~\ref{fig:p-Pb_fits_central12}.
 In the following figures $\pi$ refers to $(\pi^++\pi^-)/2$, 
$K$ refers to $K^++K^-)/2$ and similarly $p$ refers to $(p+\bar{p})/2$, $\Lambda$ refers to $(\Lambda+\bar{\Lambda})/2$,
$\Xi$ refers to $(\Xi^-+\bar{\Xi})/2$ and finally $\Omega$ refers to $(\Omega + \bar{\Omega})/2$. All fits were performed including the 
$\phi$ meson.

The
upper part of the figures compares the experimental results, shown as a black circle, to the fits obtained from the thermal
model as described in the previous section. As the logarithmic scale easily hides the quality of the fits, we also show in the middle of 
the figures the 
ratio of the experimental data to the thermal model fit. In the lowest part of the figures we show the standard deviation for each particle, calculated in the standard way as Std. Dev. 
\begin{center}
Std. Dev $\equiv$ (Experimental value -  Fit)/Error   .
\end{center}
There are no very outstanding deviations from the thermal model fits. In particular, the $\phi$ yield is reproduced  reasonably well.
\begin{figure}[htb]
%\begin{center}	
\includegraphics[trim=0cm 0cm 0cm 0cm, clip=true, width = 0.5\textwidth]{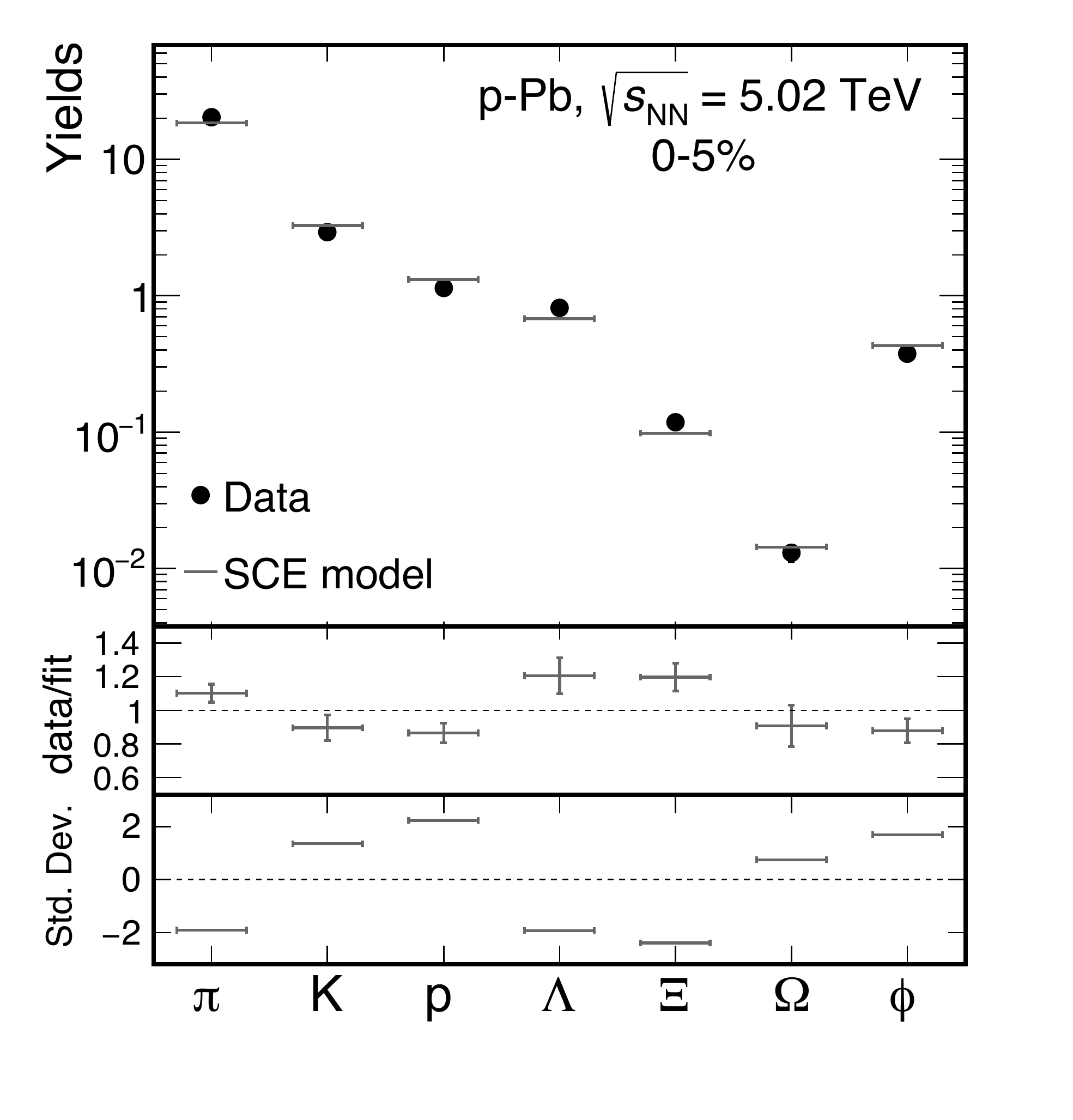} 
\includegraphics[trim=0cm 0cm 0cm 0cm, clip=true, width = 0.5\textwidth]{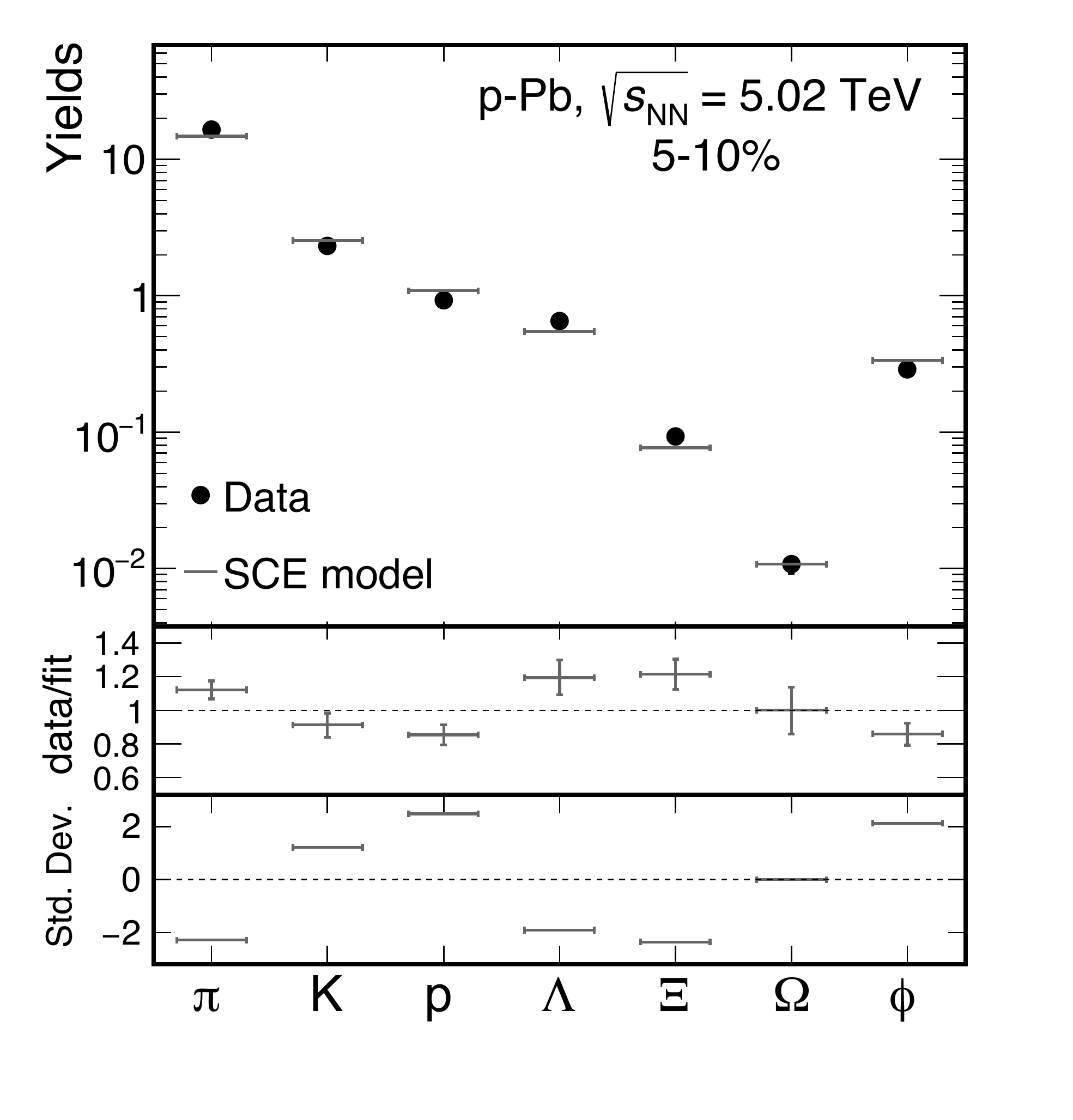} 
\caption{Comparison of data with fits  for the two  most central bins. }
\label{fig:p-Pb_fits_central12}
%\end{center}
\end{figure} 

In Fig. 2 we present our results for the two most central multiplicity bins, i.e. 0-5\% and 5-10\%. 

The pattern observed for the bins with the highest multiplicity reproduces itself for the next  two bins i.e. 10-20\%
and 20-40\% as shown in Fig. 3. 
The deviations from the 
thermal model are very similar to the previous ones. In particular, the yield of $\phi$ mesons is again reasonably well reproduced. This 
confirms the use of the strangeness non-equilibrium factor $\gamma_s$. The use of a different correlation 
radius for the strangeness suppression  would not affect the $\phi$ meson since it has net zero strangeness.

All yields are compatible with the thermal model description within two standard deviations.
\begin{figure}[htb]
%\begin{center}	
\includegraphics[trim=0cm 0cm 0cm 0cm, clip=true, width = 0.5\textwidth]{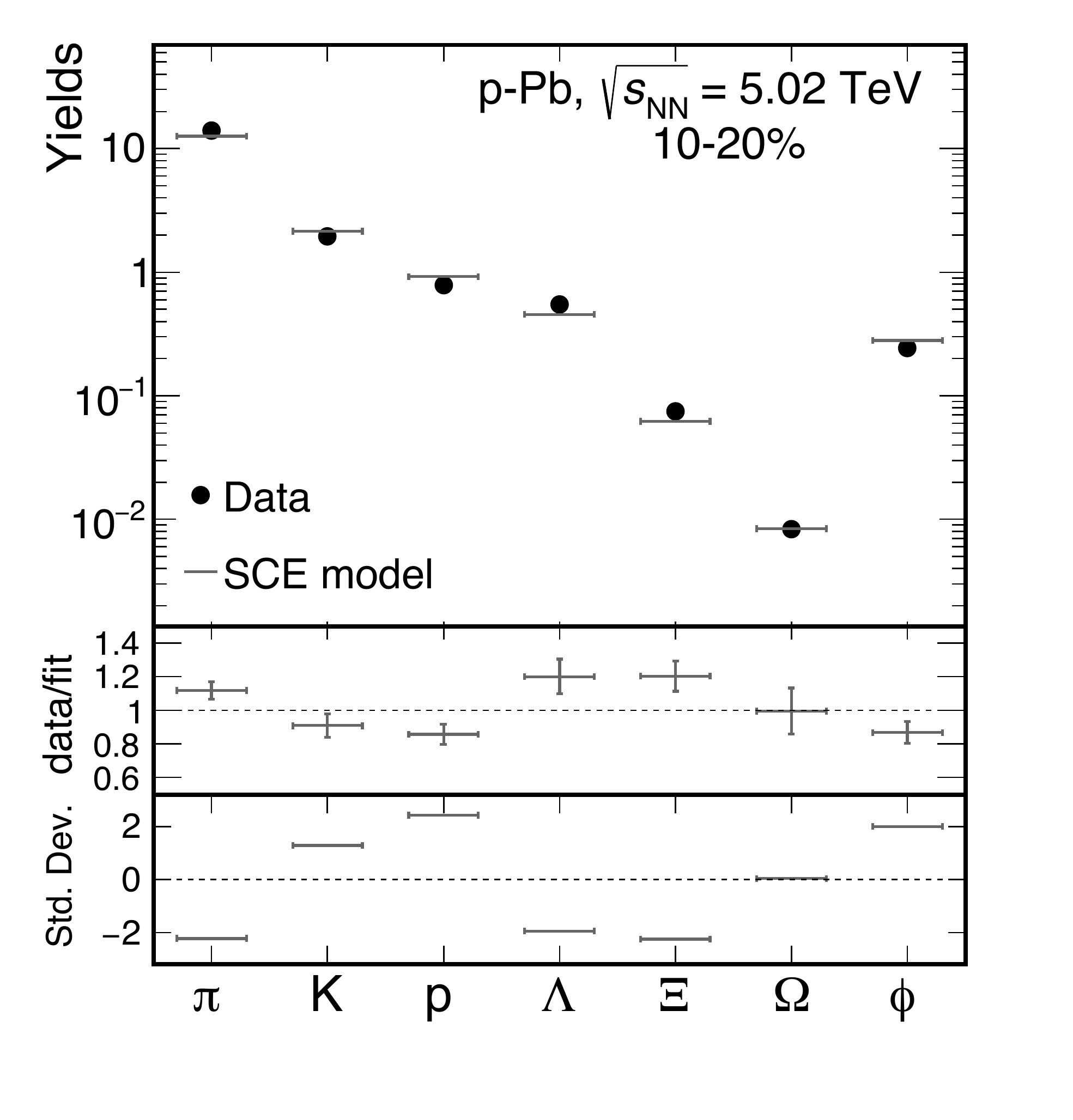}
\includegraphics[trim=0cm 0cm 0cm 0cm, clip=true, width = 0.5\textwidth]{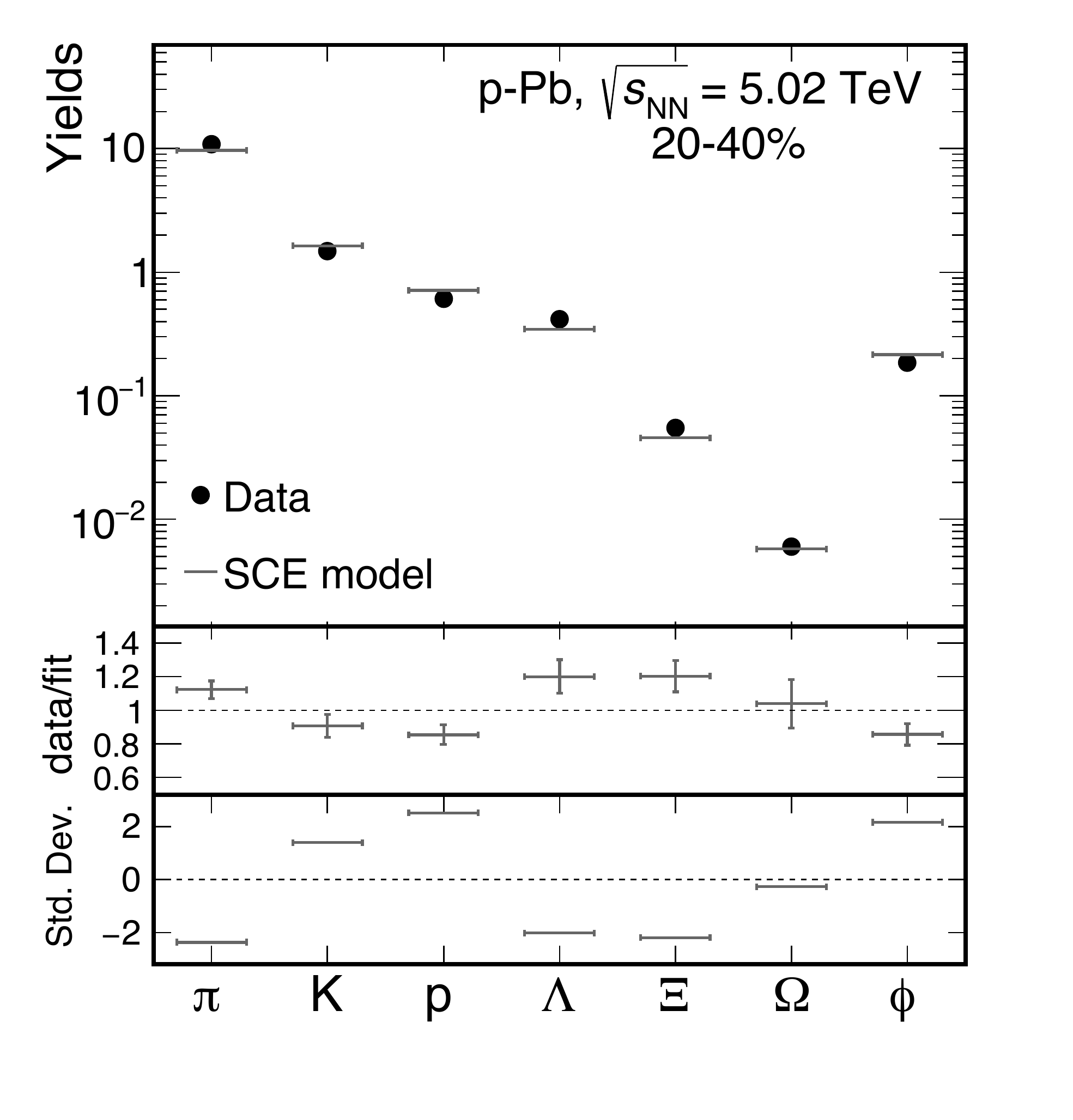} 
\caption{Comparison of data with fits  for the 10-20\% and the 20-40\% centrality bins. }
\label{fig:p-Pb_fits_central34}
%\end{center}
\end{figure} 
The next multiplicity bins are shown in Fig. 4. The left-hand side is consistent with the results shown in Figs. 2 and 3.
The right-hand pane is a peripheral one corresponding to 60-80\% centrality and shows the discrepancies with the 
thermal model fits. Especially there are clear  deviations for the $\Xi$ and $\Omega$ baryons. The $\phi$ is
reasonably well described.

\begin{figure}[htb]
%\begin{center}	
\includegraphics[trim=0cm 0cm 0cm 0cm, clip=true, width = 0.5\textwidth]{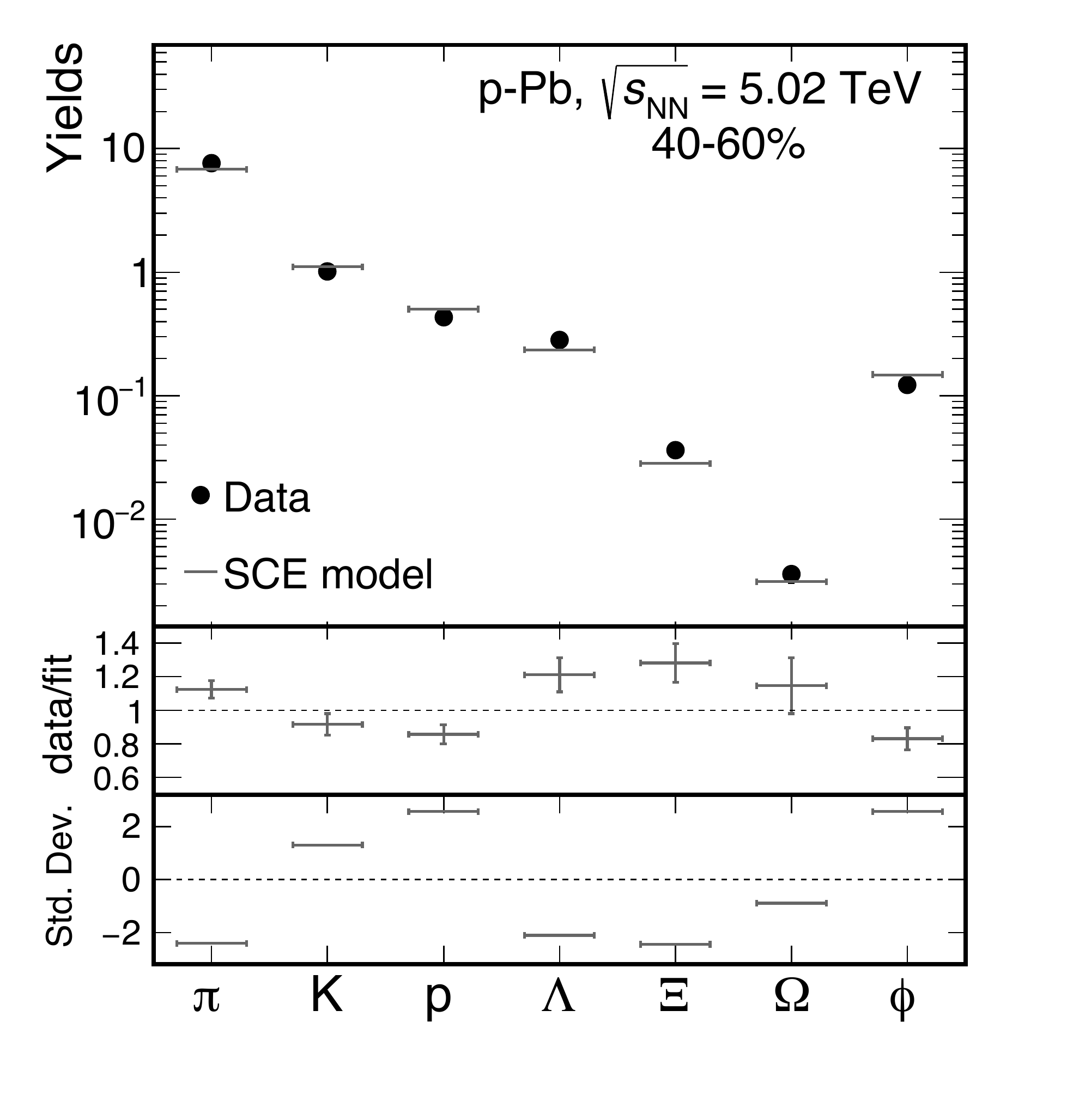} 
\includegraphics[trim=0cm 0cm 0cm 0cm, clip=true, width = 0.5\textwidth]{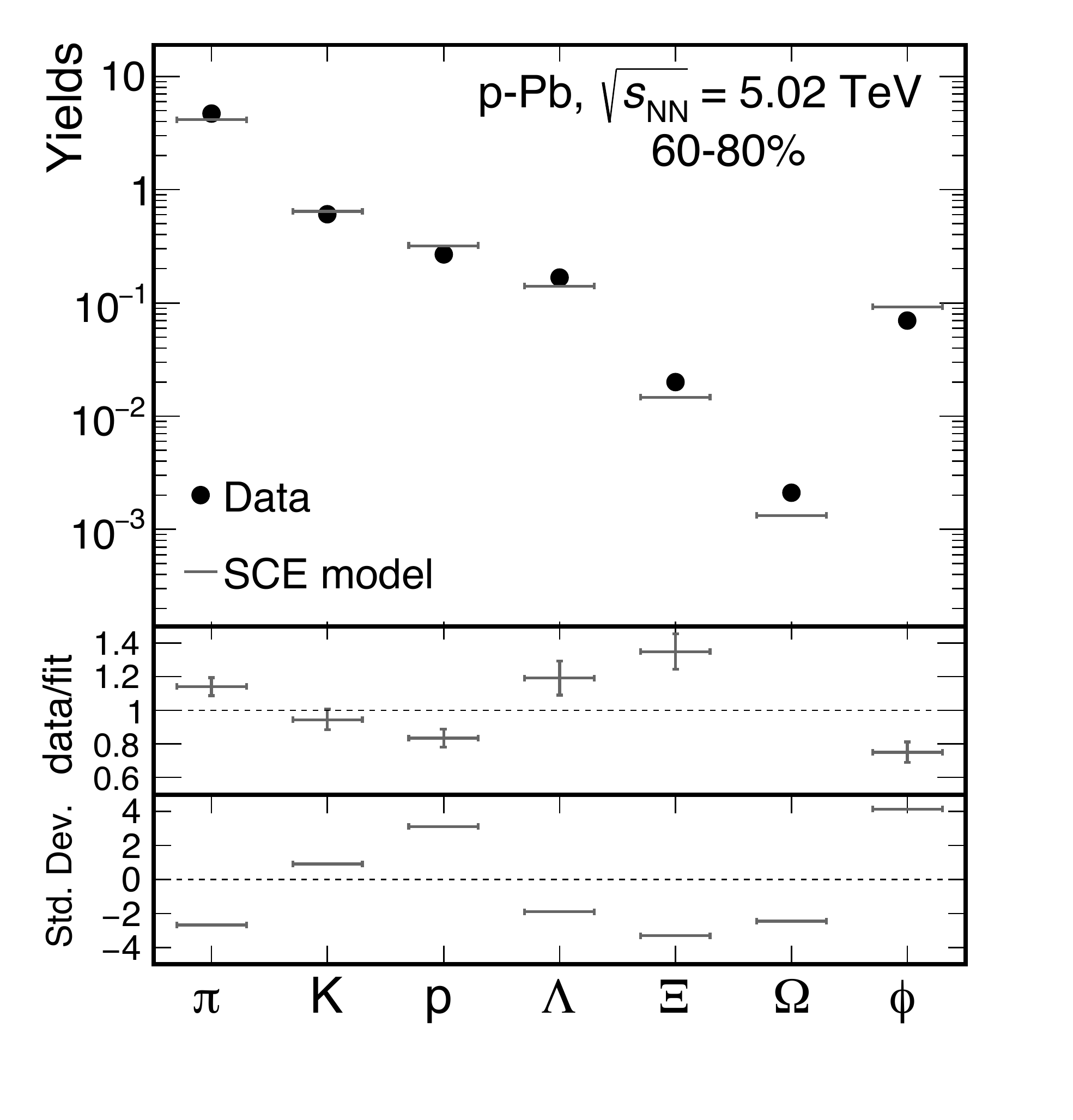} 
\caption{Comparison of data with fits for  40\% to 60\% and the 60\% to 80\% centrality bins. Note that in the centrality bin 60\% - 80\% 
the $\Omega$ has a Data/Fit ratio of 1.58  which no longer fits on the plot. }
\label{fig:p-Pb_fits_central56}
%\end{center}
\end{figure} 
The largest deviations appear for the most peripheral collisions, 80-100\%, and are shown in Fig. 5.  
Especially the values for data/fit are very high for the $\Xi$ baryon at 2.45 and for the $\Omega$ it is 4.6.  

\begin{figure}[htb]
\begin{center}	
\includegraphics[trim=0cm 0cm 0cm 0cm, clip=true, width = 0.5\textwidth]{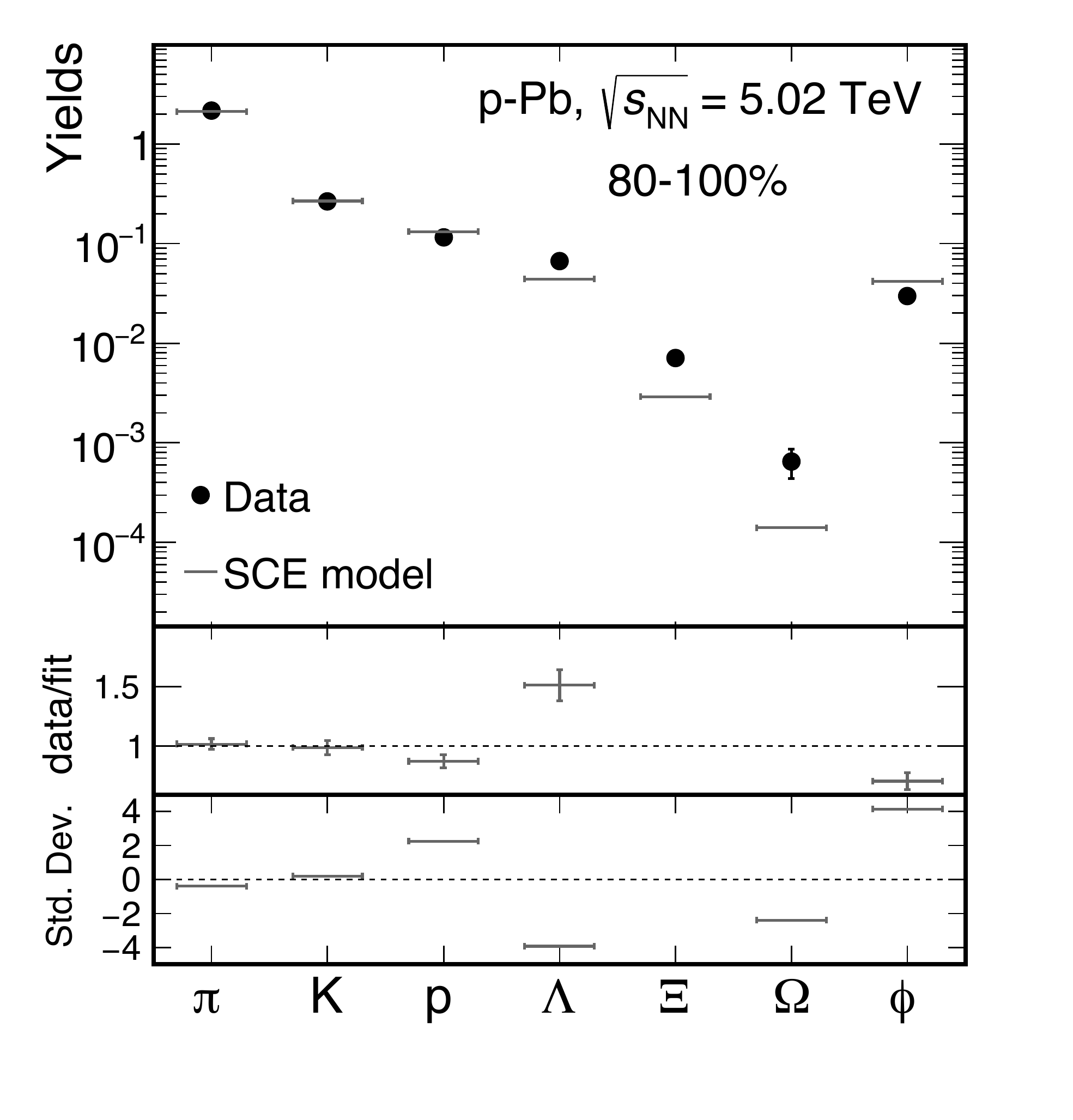} 
\caption{Comparison of data with fits for the most peripheral collisions. The lower part of the figure shows the standard deviation.
The standard deviation for the $\Xi$ is 5.5.  While the ratios data/fit for the $\Xi$ and $\Omega$ are 2.45 and 4.6 respectively.}
\label{fig:p-Pb_fits_peripheral}
\end{center}
\end{figure} 

To summarize the results obtained above we show the chemical freeze-out temperature $T_{ch}$ 
and the corresponding radius $R$ as a function of charged particle multiplicity in Fig. 6.
The values of $T_{ch}$ are remarkably independent of multiplicity except for the most peripheral bin.  
The radius determined from the yields fixes the normalization of the  yields and shows a steady increase with the 
centrality of the collision. The more particles there are in the final state, the larger the radius and, correspondingly, the volume of the system.
It should be noted that the volume increases linearly with the multiplicity. This indicates that the density of the fireball remains constant.
\begin{figure}[htb]
%	\begin{center}
%\includegraphics[trim=0cm 0cm 0cm 0cm, clip=true, width =
%0.4\textwidth, height=9cm]{CompareTch.pdf}
\includegraphics[trim=0cm 0cm 0cm 0cm, clip=true, width = 0.5\textwidth]{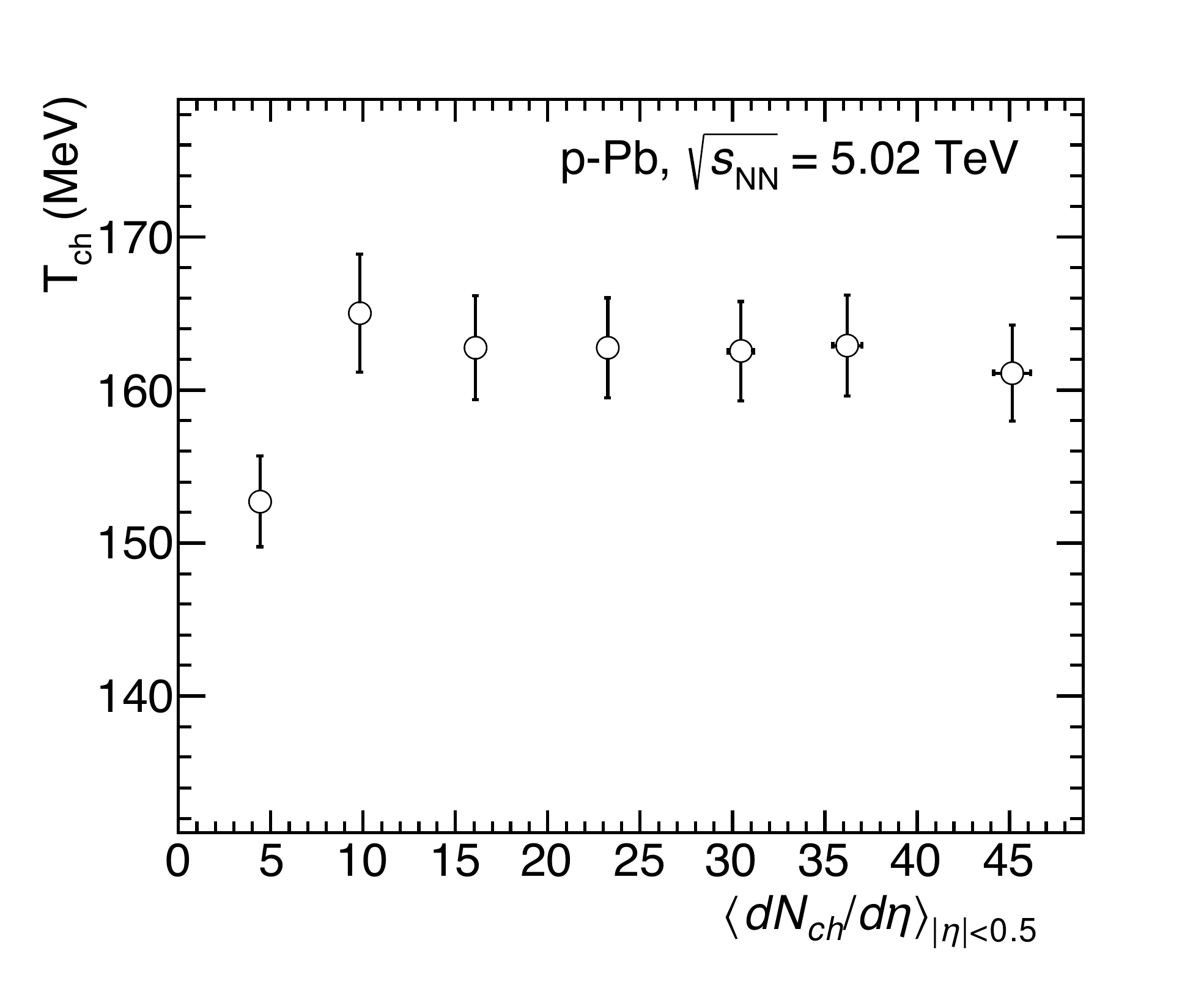}
%\includegraphics[trim=0cm 0cm 0cm 0cm, clip=true, width =
%0.4\textwidth, height=9cm]{CompareRadius.pdf}
\includegraphics[trim=0cm 0cm 0cm 0cm, clip=true, width = 0.5\textwidth]{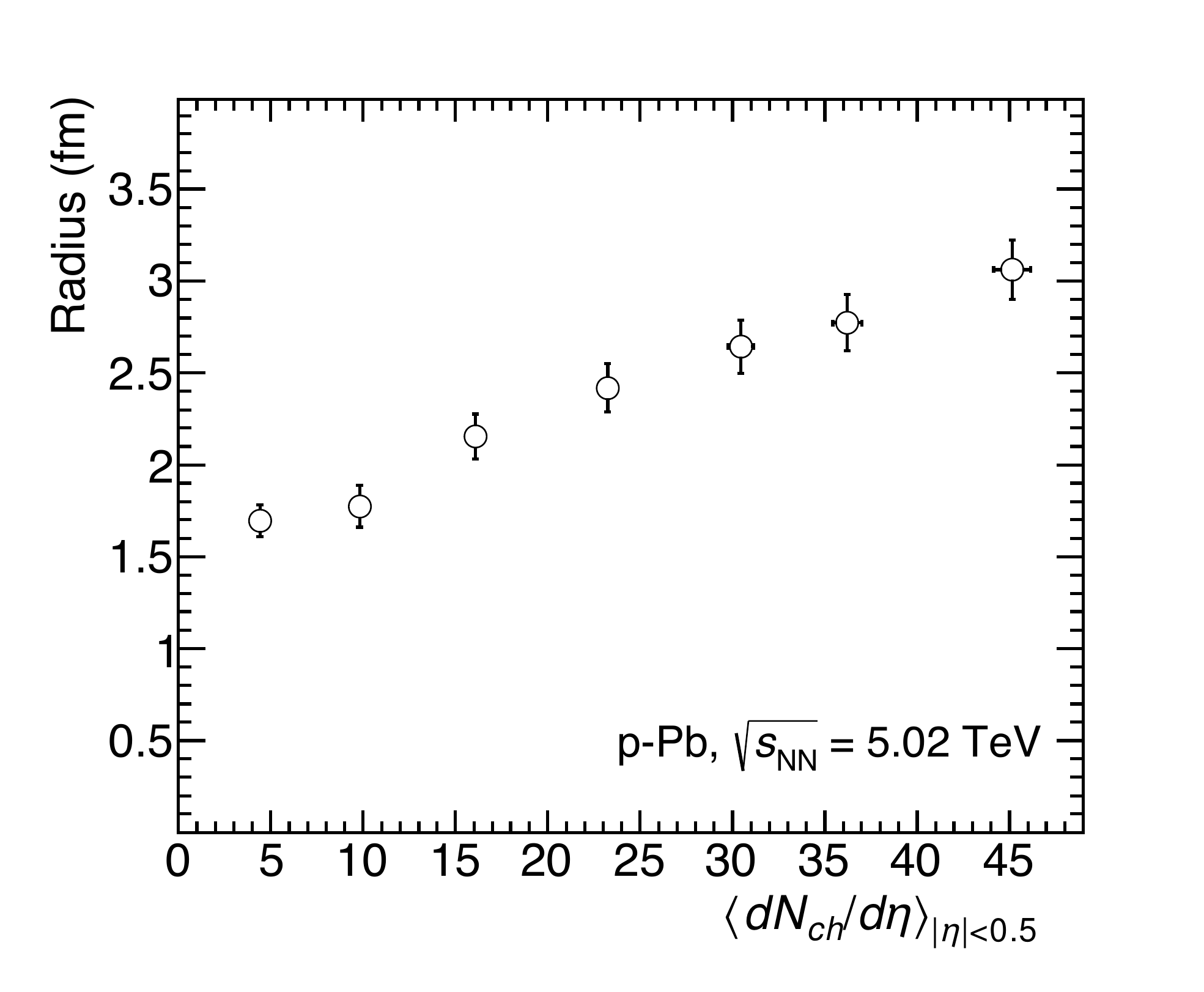}
\caption{
The  temperature (left pane) and the radius (right pane) at chemical freeze-out as a function of charged particle multiplicity
 $\langle dN_{ch}/d\eta\rangle _{|\eta|<0.5}$. 
}
\label{T_pPb}
%\end{center}
\end{figure} 
The values of the strangeness non-equilibrium factor $\gamma_s$ are shown on Fig.~\ref{gammas_pPb}.  It can be seen that $\gamma_s$ is very 
slowly increasing with multiplicity, starting from a value of about 0.9 for the most peripheral collision reaching about 0.96 for the 
bin with the highest multiplicity. This shows that the system is very close to chemical equilibrium, the largest deviation being about 10\%.
Again, being so close to chemical equilibrium is a remarkable property of the fireball produced.

Apart from the most peripheral collisions, the chemical freeze-out temperature shows a remarkable consistency for all multiplicity classes.
This shows that the final state hadrons are consistent with being  produced in a single fireball having a temperature slightly about 160 MeV.
The best fits are achieved for the most central collisions, the worst fits are for the two most peripheral bins. This is quantitatively reflected
in the values of the $\chi^2$ shown in Table I.

\begin{figure}[htb]
\begin{center}
%\includegraphics[trim=0cm 0cm 0cm 0cm, clip=true, width =
%0.6\textwidth, height=10cm]{CompareGammaS.pdf}
  \includegraphics[trim=0cm 0cm 0cm 0cm, clip=true, width = 0.6\textwidth]{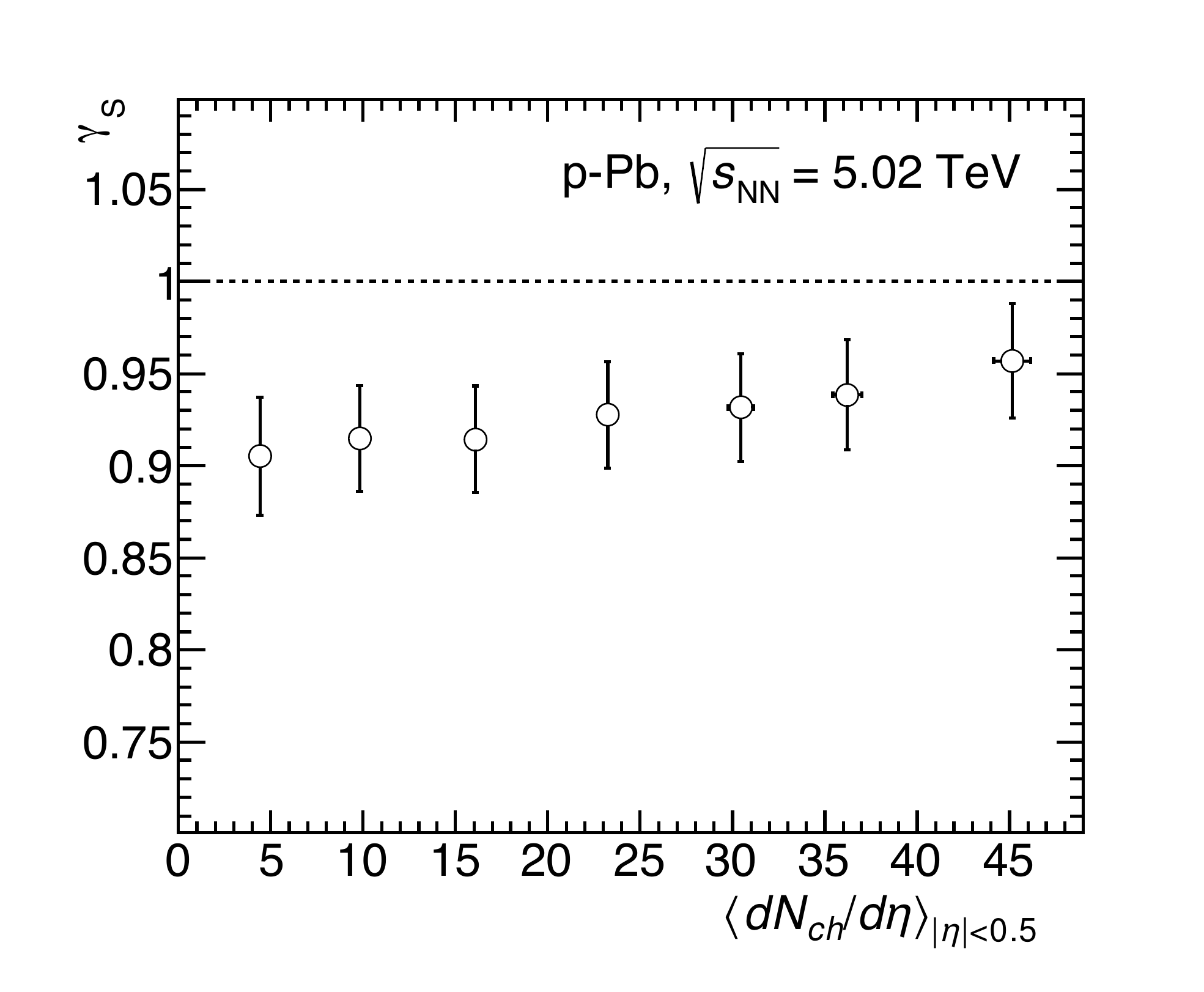}
\caption{Values of the strangeness non-equilibrium factor $\gamma_s$ as a function of the charged particle
multiplicity  $\langle dN_{ch}/d\eta\rangle _{|\eta|<0.5}$.
}
\label{gammas_pPb}
\end{center}
\end{figure} 

The information presented above is summarized  in Table I below. In addition the values of $\chi^2$ are also listed.
As can be seen by far the worst fits are obtained for the most peripheral collisions.

\begin{table*}
\begin{center}
\label{Tab:FitResults}
\begin{tabular}{|c | c | c | c | c | }
\hline
\hline
Centrality         & $T_{ch}$ (MeV) &  $R$ (fm) &  $\gamma_s$ &  $\chi^{2}/ndf$\\ 
\hline
0-5\%     &    161.10 $\pm$  3.10  &  3.06 $\pm$ 0.16   &  0.96 $\pm$  0.031  &  23.38/4  \\\hline
5-10\%    &    162.93 $\pm$  3.30  &  2.77 $\pm$ 0.15   &  0.94 $\pm$  0.030  &  26.52/4  \\ \hline
10-20\%   &    162.53 $\pm$  3.23  &  2.64 $\pm$ 0.14   &  0.93 $\pm$  0.029  &  25.44/4\\ \hline
20-40\%   &    162.75 $\pm$  3.03  &  2.42 $\pm$ 0.13   &  0.93 $\pm$  0.037  &  27.52/4\\ \hline
40-60\%   &    160.48 $\pm$  3.37  &  2.22 $\pm$ 0.12   &  0.91 $\pm$  0.029  &  20.07/4  \\ \hline
60-80\%   &    165.00 $\pm$  3.78  &  1.77 $\pm$ 0.11   &  0.91 $\pm$  0.029  &  55.40/4 \\ \hline
80-100\%  &    152.72 $\pm$  2.96  &  1.69 $\pm$ 0.09   &  0.90 $\pm$  0.032  &  74.35/4  \\ 
\hline
\hline
\end{tabular}
%}
\caption{Chemical freeze-out temperature ($T_{ch}$), radius (R), $\gamma_s$ 
and $\chi^{2}/ndf$ of the fits for various multiplicity classes. 
}
\end{center}
\end{table*}

%%%%%%%%%%%%%%%%%%%%%%%%%%%%%%%%%%%%%%%%%%%%%%%%%%%%%%%%%%%
\section{Conclusions}
%%%%%%%%%%%%%%%%%%%%%%%%%%%%%%%%%%%%%%%%%%%%%%%%%%%%%%%%%%%%%%%
The fireball being produced in $p-Pb$ collisions has simple properties, it is close to full chemical equilibrium, the temperature
is independent of the multiplicity in the final state and the radius increases smoothly which is to be expected for a system that increases
with the multiplicity in the final state.

It is remarkable that the system is always close to full chemical equilibrium with $\gamma_s$ never below
than 0.9 even for the most peripheral bins.  The chemical freeze-out temperature is also remarkably constant, except (again) for
the most peripheral bin. The radius decreases when going to more peripheral collisions, a feature which is  to be expected.

The thermal model provides a good description of the hadronic yields produced in p--Pb collisions at $\sqrt{s_{NN}}$ = 5.02 TeV.
The largest deviations occurring for the most peripheral collisions.
The chemical freeze-out temperature is independent of centrality 
$T_{ch} = 162 \pm  3$ MeV, except for the lowest multiplicity bin,  this value  is consistent with values obtained in $Pb-Pb$ collisions. 
The value of the strangeness non-equilibrium factor $\gamma_s$ is slowly increasing with multiplicity from
0.9 to 0.96, i.e. it is always very close to full chemical equilibrium. 

\vspace{1cm}
\Large{{\bf Acknowledgements}}\\
\normalsize
The authors acknowledge numerous useful discussions with J. Schukraft, B. Hippolyte and K. Redlich.
One of us (J.C.) gratefully thanks  the National Research Foundation of South Africa for financial support.
%Two of us (N.S. and L.K.) thank the Indian Research Foundation for support.
N.S. acknowledges the support of SERB Ramanujan Fellowship
(D.O. No. SB/S2/RJN- 269084/2015) of the Department of Science and Technology of India. L.K. acknowledges the support of the
SERB grant No. ECR/2016/000109 of the Department of Science and Technology of India.

%%%%%%%% Bibliography (In case of using bibtex generate the bbl requested by arXiv)
\bibliographystyle{epjc}
\bibliography{THERMAL_pPb}
\end{document}